\documentclass{vldb}

\usepackage{graphicx}
\usepackage{balance}
\usepackage[utf8]{inputenc}
\usepackage{algorithm}
\usepackage[noend]{algpseudocode}
\usepackage{algpseudocode}
\usepackage{xspace}
\usepackage[font=small]{subcaption}
\usepackage[font=bf]{caption}
\usepackage{tabularx}
\usepackage[capitalise]{cleveref}
\usepackage{biblatex}
\usepackage{xcolor,colortbl}

\bibliography{refs}
\newcommand{\bm}[1]{\mathbf{#1}}
\newcommand{\cm}[1]{\mathcal{#1}}
\newcommand{\E}{\mathbb{E}}
\newcommand{\myparagraph}[1]{\vspace{2mm} \noindent \textbf{#1}\xspace}
\newcommand{\SVT}{\textsc{SVT}}
\newcommand{\trun}{\textsc{Trunc}}
\newcommand{\ts}{\textsc{Trun}_{\textsc{SVT}}}
\newcommand{\tr}{\textsc{Trun}_{\textsc{Recur}}}
\newcommand{\rec}{\textsc{Recursive}}
\newcommand{\SQM}{\textsc{SQM}}
\newcommand{\BQM}{\textsc{BQM}}
\newcommand{\vect}{\textsc{Vectorize}}
\newcommand{\id}{\textsc{Identity}}
\newcommand{\wk}{\textsc{Workload}}
\newcommand{\itm}{\textsc{TiMM}}
\newcommand{\jtm}{\textsc{TaMM}}
\newcommand{\notrun}{\textsc{NoTrunc}}
\newcommand{\mm}{\textsc{MM}}
\newcommand{\lap}[1]{\text{Lap}(#1)}

\newcommand{\eat}[1]{}
\newcommand{\algoname}[1]{\textnormal{\textsc{#1}}}

\newcommand{\squishlist}{
	\begin{list}{$\bullet$}
		{
			\setlength{\itemsep}{0pt}
			\setlength{\parsep}{3pt}
			\setlength{\topsep}{3pt}
			\setlength{\partopsep}{0pt}
			\setlength{\leftmargin}{1.5em}
			\setlength{\labelwidth}{1em}
			\setlength{\labelsep}{0.5em} } }

	\newcommand{\squishend}{
\end{list}  }

\newcolumntype{s}{>{\hsize=0.5\hsize}X}
\newcolumntype{m}{>{\hsize=1.5\hsize}X}

\newtheorem{example}{Example}

\newtheorem{theorem}{Theorem}

\vldbTitle{MM \& truncation}
\vldbAuthors{Yikai Wu, David Pujol, Ios Kotsogiannis, and Ashwin Machanavajjhala }
\vldbDOI{https://doi.org/10.14778/xxxxxxx.xxxxxxx}
\vldbVolume{xxxx}
\vldbNumber{xxx}
\vldbYear{xxxx}

\makeatletter
\def\@copyrightspace{\relax}
\def\@mkbibcitation{\relax}
\makeatother

\begin{document}

\title{Answering Summation Queries for Numerical Attributes under Differential Privacy}

\numberofauthors{1}

\author{
	\alignauthor Yikai Wu, David Pujol, Ios Kotsogiannis, Ashwin Machanavajjhala\\
	\affaddr{Duke University}\\
	\email{{\small yikai.wu@duke.edu \{dpujol, iosk, ashwin\}@cs.duke.edu}}
}


\maketitle
\begin{abstract}
    In this work we explore the problem of answering a set of sum queries under Differential Privacy. This is a little understood, non-trivial problem especially in the case of numerical domains. We show that traditional techniques from the literature are not always the best choice and a more rigorous approach is necessary to develop low error algorithms.
\end{abstract}
\section{Introduction}\label{sec:intro}

In recent years, Differential Privacy (DP) \cite{DiffPriv} has emerged as the de-facto privacy standard for sensitive data analysis. Informally, the output of a DP mechanism does not significantly change under the presence or absence of any single tuple in the dataset. The privacy loss is captured by the privacy parameter $\epsilon$, also referred to as the \textit{privacy budget}. 
DP algorithms usually work by adding noise to query results. This noise is calibrated to the  parameter $\epsilon$ and the \textit{query sensitivity}, i.e., the maximum change in the query upon the deletion/addition of a single row in the dataset.


In this work we focus on privately releasing sum queries over numerical attributes. A sum query involves adding the values of all records meeting certain criteria. These queries can be a powerful tool for data analysts to to gain insight about a dataset. In \cref{ex:salaries} we present a use case for sum queries.

\begin{figure}[h]
    \centering
    \includegraphics[width=0.5\linewidth]{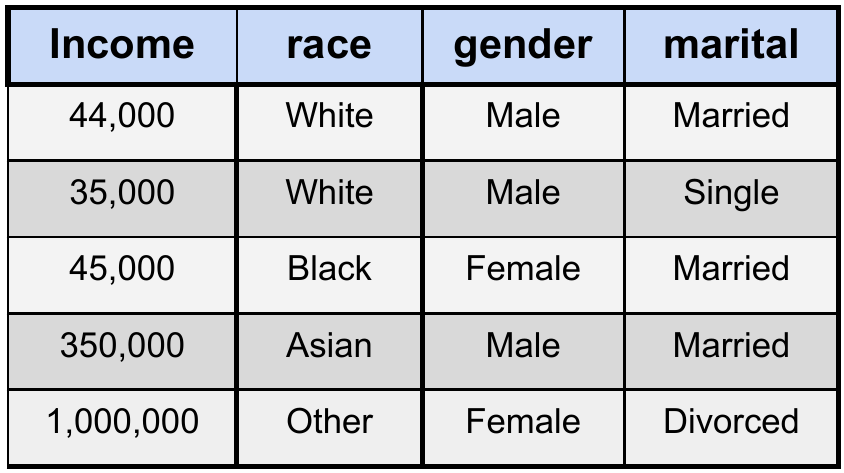}
    \caption{Sample of a population statistics dataset.}
    \label{fig:db}
\end{figure}

\begin{example}\label{ex:salaries}
Consider  the dataset of \cref{fig:db} and a social scientist querying it for insight on income distributions. More specifically, the scientist queries the database for the sum of salaries of people with income less than \$30k, \$40k, \$50k up to \$1M, in which case and the answers are: $0, 35k, \allowbreak 124k,$ and $1.474M$ respectively.
\end{example}

Answering a single sum query under DP proves to be challenging due to their high sensitivity -- the addition/removal of a single row can have a dramatic effect on the query. In \cref{ex:salaries} we see that the the sensitivity of the final sum query is $10^6$. 
One way to reduce this sensitivity is via the addition of a truncation operator, which truncates the queries such that any value above a certain threshold $\theta$ only contributes $\theta$ to the query answer.
However, truncation techniques introduce bias to the final answer, even in the absence of any noise mechanism.

In this work we focus on answering a batch of sum queries under a common privacy budget.
This problem is challenging for two reasons. First, prior work on batch query answering, such as matrix mechanism, $\id$, and $\wk$\cite{DPBench,Matrix}, is focused on workloads of queries with similar sensitivities. For example, $\wk$ applies noise proportional to the query with the worst sensitivity to all of the queries.  

Second, although post-processing techniques from prior work \cite{Hay2010} has shown the ability to dramatically reduce the final error for a workload of queries, it is not clear how such techniques would fare for post-processing noisy answers each of which having a different bias (e.g., noisy answers have different truncation thresholds). 

Our main contributions are as follows:
\squishlist
    \item We introduce methods of implementing truncation on sum queries effectively reducing their sensitivity.
    
    \item In \cref{sec:batch_queries}, we propose 2 new DP algorithms ($\itm$ and $\jtm$) for answering batch of sum queries under the same privacy budget.
    
    \item In \cref{sec:exp} we conduct a study on a U.S. Census dataset where we: (a) highlight the importance of truncation for sum queries and (b) evaluate the performance of our proposed algorithms with $\jtm$ offering the overall best performance.
    
    \item We explore the effects of post-processing for noisy answers that are heterogeneously biased.
\squishend

\section{Background}\label{sec:background}
\eat{\myparagraph{Differential Privacy}
(DP) \cite{DiffPriv}. Informally, the output of a DP mechanism does not significantly change under the presence or absence of any single tuple in the dataset. The privacy loss is captured by $\epsilon$ the privacy parameter, often referred to as the \textit{privacy budget}. 
Additionally, under DP the privacy loss gracefully degrades under composition \cite{dwork2014:textbook}:  for mechanisms $M_1, \dots  M_k$ each satisfying $\epsilon_i$-DP their sequential execution satisfies  $\sum_{i=1}^k\epsilon_i$-DP.
}

\myparagraph{Data Representation}
We consider databases where each tuple corresponds to a single individual and have a single numerical attribute. More specifically, let $D$ be a multiset of records drawn from a numerical domain $\mathcal{A} =  \{a_1 , a_2 \dots \}$. For instance, the database of \Cref{ex:salaries} consists of records drawn from $\mathbb{N}$, the  domain of natural numbers. 

Many differentially private algorithms use the \textit{vector form} of a database: $\mathbf{x}^D$. More specifically, given a set of buckets $\mathcal{B} = \{(l_1, u_1), \dotsc, (l_k, u_k) \}$ the original set of records $D$, is transformed to a vector of counts $\mathbf{x}^D$, where $x^D_i$ is the number of individuals in $D$ with value $t \in [l_1, u_1]$. For brevity, in the remainder we use the simplified notation $\mathbf{x}$ to refer to the vector form of database instance $D$.

\myparagraph{Prefix Sums}
As noted earlier, in this work we focus on summation queries over numerical attributes. More specifically, for database instance $D$ we consider the \textit{prefix sum} query defined as
$q_i(D) = \sum_{t \in D, t \leq i} t $. 

The query $q_i(D)$ returns the sum of values of tuples in $D$ with value less or equal to $i$. We also consider sets of prefix sum queries, where each query $q_i \in Q$ has an increasing threshold $i$. In the motivating \cref{ex:salaries}, the query ``Total cost to company for employees with salary at most $30k$'' is encoded by the prefix query $q_{30k}$ and the full workload of queries asked is encoded by $Q = \{q_{30k}, q_{40k},q_{50k}, \dotsc, q_{1M}\}$. Prefix sum queries can be vectorized to 0-1 lower triangular matrix $\bm{W}$, we call it workload matrix. 

Sum queries on the vector form require every bucket  to have a weight. We can define the weight of the bucket $(l_i, u_i)$ to be $u_i$. Thus, we can define the weight matrix $\bm{D}$ as a diagonal matrix with $\bm{D}_{ii} = u_i$. The true answer is thus $\bm{WDx}$, and we call $\bm{WD}$ weighted workload matrix.\par

\myparagraph{Sparse Vector Technique}
$(\SVT)$ \cite{dwork2014:textbook} is a differentially private mechanism which reports if a sequence of queries  lie above a chosen threshold. $\SVT$ takes in as input a sequence of queries $\{f_i\}$ and a threshold $T$. $\SVT$ then outputs the first query which lies above the threshold.

\myparagraph{Recursive Mechanism}
 \cite{Chen13:recursive} is an algorithm for answering monotone SQL-like counting queries of high sensitivity. It internally finds a threshold $\hat{\Delta}$ to reduce the sensitivity of the query 
and then constructs a recursive sequence of lower sensitivity queries which can be used to approximate the input query. The parameter $\hat{\Delta}$ trades-off bias for variance. 

\myparagraph{Matrix Mechanism}
 (\mm) \cite{Li15matrix} is a differentially private mechanism that allows for the private answering of batch queries. $\mm$ takes in as input a workload of queries $\bm{W}$ in matrix form and a database $\bm{x}$ in vector form. It computes a differentially private answer to $\bm{Wx}$ by measuring a different set of strategy queries $\bm{A}$ and reconstructing answers to $\bm{W}$ from noisy answers to $\bm{A}$. 

\myparagraph{Identity and Workload}
\cite{DPBench,Zhang:2018:EFD:3183713.3196921} are particular strategies in the $\mm$ framework. $\id$ simply adds noise to each count in $\bm{x}$ and then computes the query workload as normal. $\wk$ however first computes the true query answers $\bm{Wx}$ then adds noise to the true answers based off the query with the highest sensitivity.

\section{Answering Sum Queries}\label{sec:sum_queries}
We now introduce the methods developed for answering sum queries. In \cref{sec:single_query} we discuss answering a single sum query and in \cref{sec:batch_queries} we propose algorithms for answering a workload of sum queries.

\subsection{Answering a Single Query}\label{sec:single_query}
A sum query can have very large (and even unbounded) sensitivity, which often leads to prohibitively large scale of injected noise for satisfying the privacy guarantee. One simple, yet effective, method to reduce the sensitivity is by truncating the values of tuples in the original database before answering the sum query. For a database $D$, and a threshold $ \theta $, all tuples of $ D $ with value higher than $ \theta $ are replaced with the value $ \theta $. More specifically, let $ \textsc{Trunc}_\theta$ be a truncation operation on queries that is defined as follows: $ \textsc{Trunc}_\theta(q_i)(D) = \sum_{t\in D, t \leq i} \min(t, \theta)$. Then, for any sum query $ q_i $, $\textsc{Trunc}_\theta(q_i)$  will have sensitivity $\min(i, \theta)$. In other words, asking a truncated sum query can possibly have smaller sensitivity.

This sensitivity reduction comes with a cost in bias since truncation reduces the true answer of $q_i$ even in the absence of any noise mechanism. Thus, the choice of $\theta$ is crucial, since very small values (e.g., $\theta = 1$) which lead to low sensitivity values, also lead to an increased bias. For example if we use $\theta=1$ as a truncation threshold for $\cref{ex:salaries}$ we find that the answers to $q_{30k},q_{50k}$ and $q_{1M}$ are all $5$.  Similarly, large values of truncation will have small bias, but might not decrease the sensitivity. 

At the same time, any choices of $\theta$ need to be done either (a) data independently (e.g., using an oracle), or (b) using the sensitive data under differential privacy. In the following we explore $2$ methods of \textit{privately choosing} a threshold $\theta$.

\myparagraph{Recursive Mechanism}
The recursive mechanism can be used to privately find a truncation threshold. We simply then use $\theta = \hat{\Delta}$ as the truncation threshold. The results would be equivalent to implement the whole recursive mechanism on the sum query. The proof and details of the algorithm are in the Appendix.

\myparagraph{Sparse Vector Technique} Likewise we can use $\SVT$ to chose  $\theta$. We choose the sequence of functions $\{f_{u_1}, f_{u_2}, \ldots\}$, where $f_{u_i}$ is a counting query which counts the number of tuples with weight at most $u_i$. We choose a ratio $r$ of the database which we would like to not be truncated. We then set the $\SVT$ threshold  $T= rN$. $\SVT$ will then return $u_k$ where $f_{u_k}$ is the first query  where the number of tuples less than $u_k$ is greater than $rN$. We then use $\theta = u_k$ as our truncation threshold.
When choosing the sequence of counting queries we begin with $f_s$ where $u_1 = s$ is a number far below the expected $\theta$. We then let the sequence $\{u\} = \{u_1, u_2, \ldots\}$ be a strictly increasing sequence. We found that linearly increasing $\{u\}$ with a reasonable interval gives a very slow performance and returns a smaller $\theta$ than expected. As such we use an exponential increase in $\{u\}$. We thus define the sequence of $\{u\}$ with a parameter $c$ such that $u_i = sc^{i-1}$.



\begin{table*}[hbt]
\begin{tabularx}{\linewidth}{|s|m|}
\hline
Algorithm                & Description \\ \hline
$\SQM$\newline(Single Query Mode) & Split the budget equally for each single query. Select truncation threshold independently for each query. Add noise according to the query bound and truncation threshold. \newline Output: $\trun_\theta(q_i)(D) + \text{Lap}[\min(\theta, i)/\epsilon_2]$\\ \hline
$\id$\newline(Identity Strategy) & Generate truncated weight matrix $\bm{T}$. Add Laplace noise of scale 1 to the vector form $\bm{x}$. Perform truncated query $\bm{WT}$ on the noisy counts.\newline Output: $\bm{WT}[\bm{x}+(1/\epsilon_2)\tilde{\bm{b}}] $ \\ \hline
$\wk$\newline(Workload Mechanism) & Generate truncated weight matrix $\bm{T}$. Perform query $\bm{W}$ on the weighted vector form $\bm{Tx}$ and add Laplace noise with the scale of the L-1 norm (maximum column norm)  of $\bm{WT}$.\newline Output:$\bm{WTx}+(\|\bm{WT}\|_1/\epsilon_2)\tilde{\bm{b}}$       \\ \hline
$\itm$\newline(Truncation-independent Matrix Mechanism) & Generate truncated weight matrix $\bm{T}$. Select a strategy matrix $\bm{A}$ based on workload matrix $\bm{W}$. We then perform matrix mechanism on the weighted vector form $\bm{Tx}$.\newline Output: $\bm{W}[\bm{Tx} + (\|\bm{AT}\|_1/\epsilon_2)\bm{A^+\tilde{b}}]$    \\ \hline
$\jtm$\newline(Truncation-aware Matrix Mechanism) & Generate truncated weight matrix $\bm{T}$. Select a strategy matrix $\bm{A}$ based on weighted workload matrix $\bm{WT}$. We then perform matrix mechanism on the vector form $\bm{x}$.\newline Output:  $\bm{WT}[\bm{x}+(\|\bm{A}\|_1/\epsilon_2)\bm{A^+\tilde{b}}]$  \\ \hline
\end{tabularx}
\caption{Algorithms for answering a workload of sum queries.}\label{tab:algos}
\label{tb:alg}
\end{table*}

\subsection{Answering a Workload of Queries}\label{sec:batch_queries}
We now propose a general routine for answering a workload of sum queries. In \cref{tb:alg} we offer  descriptions of 3 baseline algorithms and 2 new algorithms. In the table, we define $\bm{\tilde{b}}$ as a vector of i.i.d. random variables drawn from a Laplace distribution with mean 0 and scale 1.

\myparagraph{Baseline Algorithms}
    The first baseline $\SQM$, which naively splits the budget across all queries and sequentially answer them with the Laplace mechanism. Additionally, direct implementations of $\id$, $\wk$ and Matrix Mechanism ($\mm$) are also considered. Note that sum queries can have largely different sensitivities for queries in a query set and very high sensitivity for queries involving tuples with large weights. In \cref{ex:salaries}, there may be only a few queries involving individuals with very high salaries like 1 million. To ensure the privacy of these individuals, we need to add a very large noise to all the queries if we use $\wk$ and some queries if we use $\id$ and $\mm$. This characteristic of sum query sets makes these algorithms sub-optimal. Thus, we propose truncated versions of these algorithms.
    
\myparagraph{Truncation}
    As  in  the  case  of  answering  a  single  query, truncation  is  a  useful  technique  to  reduce  the  sensitivity of  sum  queries. We split the privacy budget with a parameter $\rho$ to assign a private budget $\epsilon_1 = \rho\epsilon$ to $\SVT$ or $\rec$ to find a truncation threshold $\theta$. We  then obtain a truncated weight matrix $\bm{T}$ by changing all the values in $\bm{D}$ larger than $\theta$ to $\theta$. Thus, when we use $\bm{WT}$ as the weighted workload matrix, it is equivalent to applying $\textsc{Trunc}_\theta$ to every query of $\bm{W}$. We call this truncation method $\ts$ if $\SVT$ is used and $\tr$ if $\rec$ is used. We can thus use $\bm{T}$ instead $\bm{D}$ as the weight matrix to implement $\id$ and $\wk$.\par 
\myparagraph{Truncated Matrix Mechanisms}
    As discussed earlier, $\mm$ and HDMM\cite{HDMM} is preferred  for answering a batch of queries since it optimizes for the input workload $\mathbf{W}$ using a strategy matrix $\mathbf{A}$.
    Ideally, in our problem, we want to jointly optimize the strategy matrix $\bm{A}$ and the truncated weight matrix $\bm{T}$ w.r.t. to the workload and the data. 
    In addition, $\bm{T}$ can be any matrix instead of a diagonal matrix obtained from $\bm{D}$ with a numerical threshold. We now introduce 2 heuristic algorithms to implement $\mm$ with truncation.

Both algorithms obtain $\bm{T}$ using $\ts$ or $\tr$ as described above.  After obtaining $\bm{T}$, one way is to optimize $\bm{A}$ using the workload matrix $\bm{W}$ without using the results of truncation $\bm{T}$. We call this Truncation-independent Matrix Mechanism (\itm). A different approach is to optimize $\bm{A}$ using the weighted workload matrix $\bm{WT}$. We call this Truncation-aware Matrix Mechanism (\jtm). Both procedures are described in \cref{tb:alg}. The full description of the algorithm and the analytical expressions for expected error are presented in the appendix. 
\section{Experiments}\label{sec:exp}
We experimentally evaluate our algorithms on a U.S. Census dataset, our key findings are: (a) truncation improves the error of each algorithm tested, (b) our proposed algorithm $\jtm$ performs the best, and (c) traditional post-processing techniques do not necessarily reduce the error.

\myparagraph{Dataset}
We use CPS, a dataset derived from the publicly available Current Population Survey \cite{Zhang:2018:EFD:3183713.3196921}. More specifically, CPS contains over $40k$ tuples corresponding to individuals and 4 attributes: income, race, gender, and marital status. We project on the income attribute to derive our dataset.

\myparagraph{Queries}
We evaluate using $3$ workloads of prefix sum queries: $Q_1$, $Q_2$, and $Q_3$ containing $1000$, $100$, and $10$ queries respectively. More specifically, $Q_1=\{q_{\sigma_1}, q_{\sigma_2}, \ldots, q_{\sigma_{1000}}\}$, for $\sigma_i$ = $800i$; $Q_2 = \{q_{\sigma_{10}}, q_{\sigma_{20}}, \ldots, q_{\sigma_{1000}}\}$; and  $Q_3 = \{q_{\sigma_{100}}, q_{\sigma_{200}}, \ldots, q_{\sigma_{1000}}\}$.

\myparagraph{Vectorization}
All $\BQM$ algorithms presented in \cref{sec:batch_queries} operate on the vector form of the dataset and workload. We vectorize our dataset and queries using the set of bins $\mathcal{B} = \{(\sigma_{i-1} -1, \sigma_i)\}_{i \in [1000]}$. The workloads $Q_1, Q_2$, and $Q_3$ are also vectorized to $\mathbf{W}_1, \mathbf{W}_2,$ and $\mathbf{W}_3$ respectively. Where $\mathbf{W}_1$ is a $1000 \times 1000$ lower triangular matrix, $\bm{W}_2$ is a $100\times 1000$ matrix, and $\bm{W}_3$ is a $10\times 1000$ matrix.



\myparagraph{Algorithms}
We experimented on all 5 algorithms listed in \cref{tab:algos}. 
Each algorithm is executed without the \textsc{Trunc} subroutine ($\notrun$), or with the \textsc{Trunc} subroutine with a threshold learned using \SVT ($\ts$), or $\rec$ ($\tr$) for a total of 15 different configurations. We ran each algorithm on a unique input for a total of 100 independent trials and we report aggregate statistics.

Algorithms using the $\textsc{Trunc}_\theta$ subroutine, use a budget split $\rho \in \{0.1, 0.5\}$ to learn $\theta$ privately; for brevity in our results we only report for $\rho = 0.1$
For the  $\SVT$ subroutine we chose $r =0.998, c=1.2,$ and $s=5\times 10^4$. $\rec$ is implemented with $\beta = 2\epsilon_2/5, \theta =5 \times 10^4, \mu = 0.5$.
We ran our experiments on the environment of ektelo\cite{Zhang:2018:EFD:3183713.3196921}. We used the \textsc{GreedyH} algorithm provided by ektelo to compute the strategy matrix $\bm{A}$.
Since our workload is prefix sums, we used isotonic regression as a post-processing step for all results we present -- unless explicitly stated. 
Across all algorithms, we fixed the privacy parameter to $\epsilon = 0.01$.

\myparagraph{Error}
For  query $q_i$ and algorithm $\mathcal{A}$, we report the \textit{relative error} of $\mathcal{A}$  on $q_i$, defined as follows: $\text{Error}_\mathcal{A}(q_i) = \frac{|\hat{y}_i - y_i|}{\max(y_i, \delta)}$ where $\hat{y}_i$ is the noisy answer of $q_i$, $y_i$ is the true answer, and $\delta$ is positive parameter -- in all experiments we use $\delta = 100$.

\begin{figure}[t]
      \begin{subfigure}[b]{0.5\linewidth}
        \centering
        \includegraphics[width=1\linewidth]{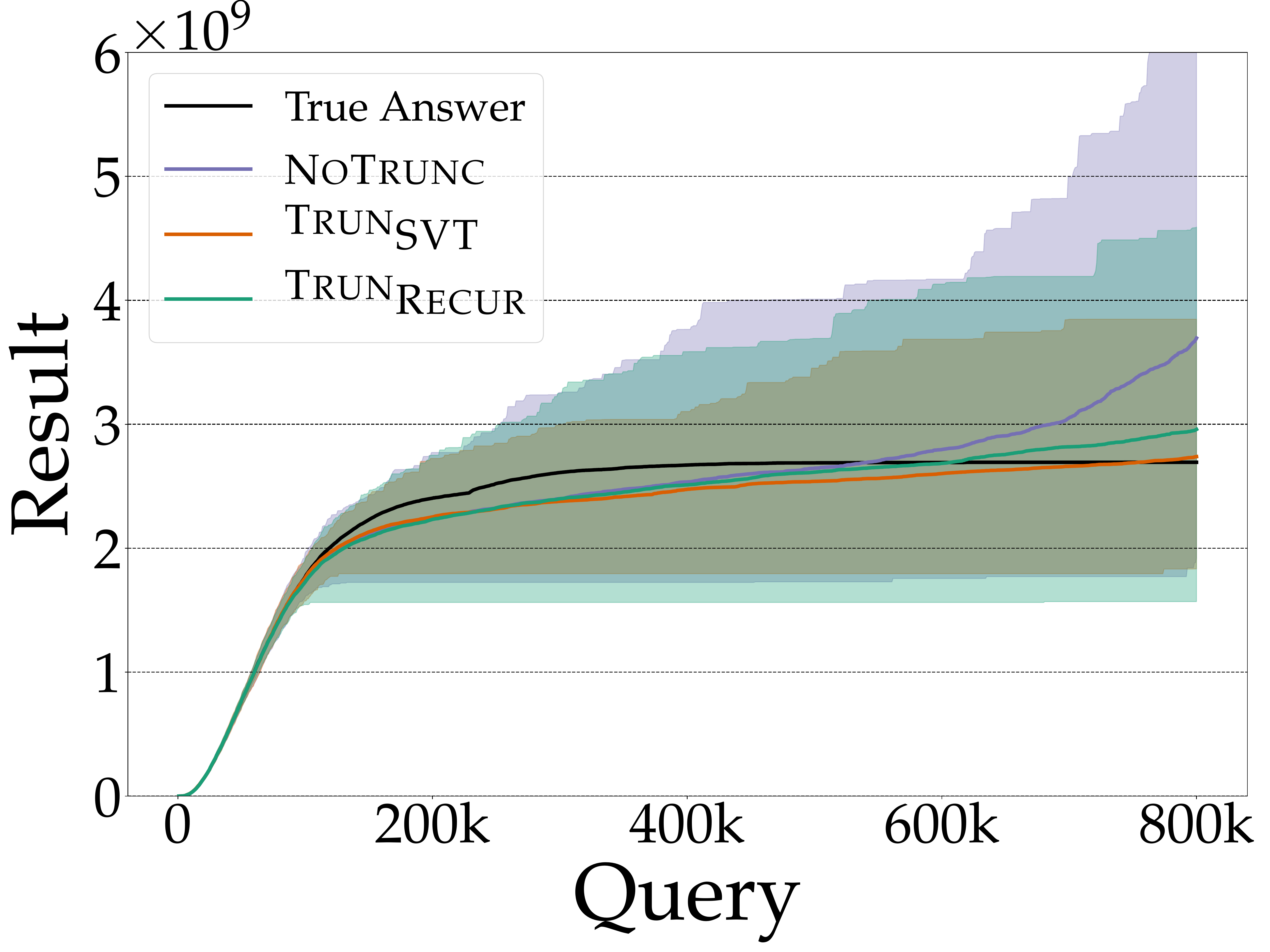}
        \caption{\small \algoname{Identity} using \textsc{Trunc}}
        \label{fig:Identity}
      \end{subfigure}
      \begin{subfigure}[b]{0.5\linewidth}
        \centering
        \includegraphics[width=1\linewidth]{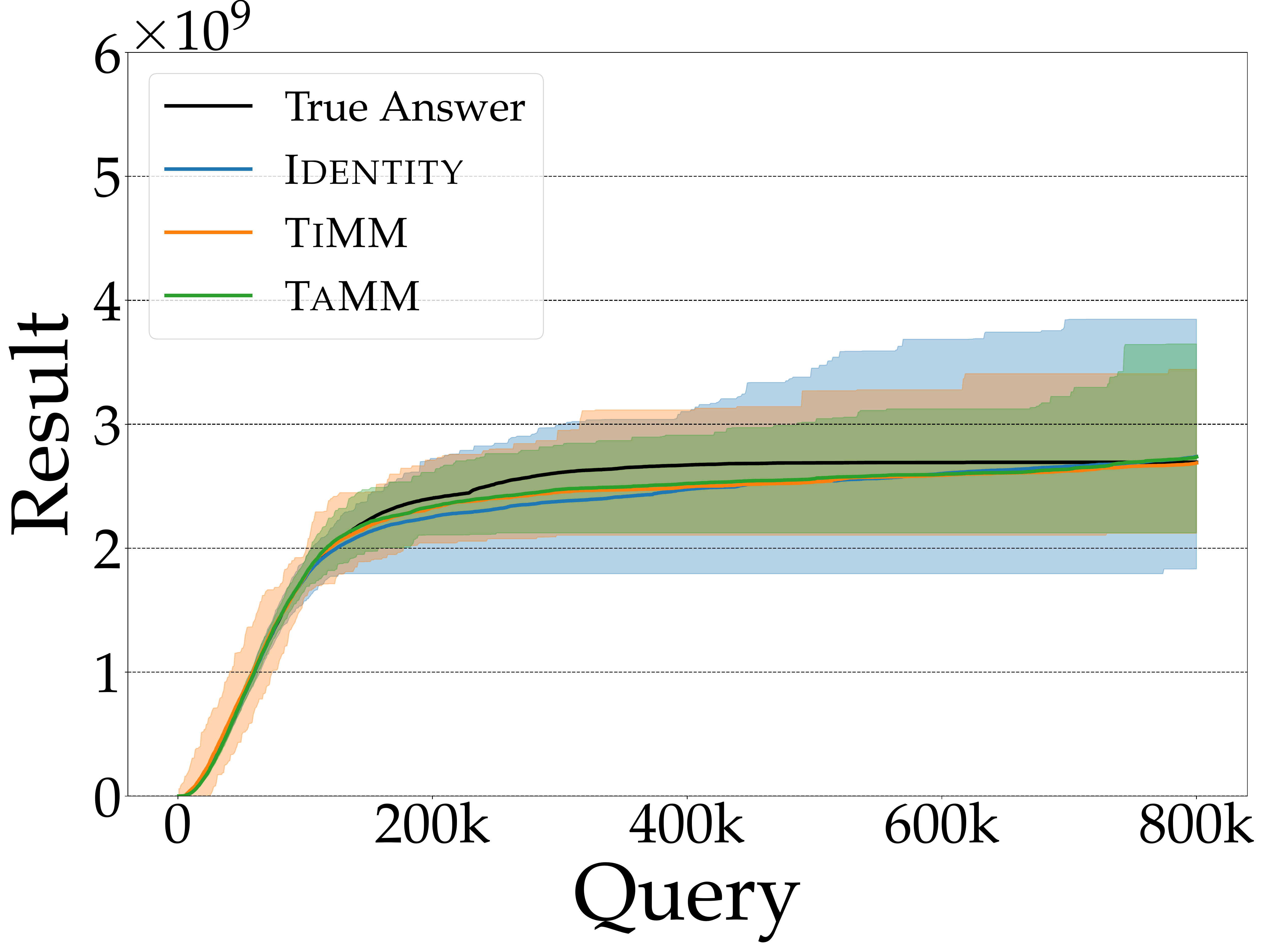}
        \caption{$\mm$ using $\ts$}
        \label{fig:SVT}
      \end{subfigure}
      \caption{Noisy answers of DP algorithms for $Q_1$}
      \label{fig:result}
\end{figure}

\begin{figure}[t]
    \begin{subfigure}[b]{0.5\linewidth}
    \centering
        \includegraphics[width=1\linewidth]{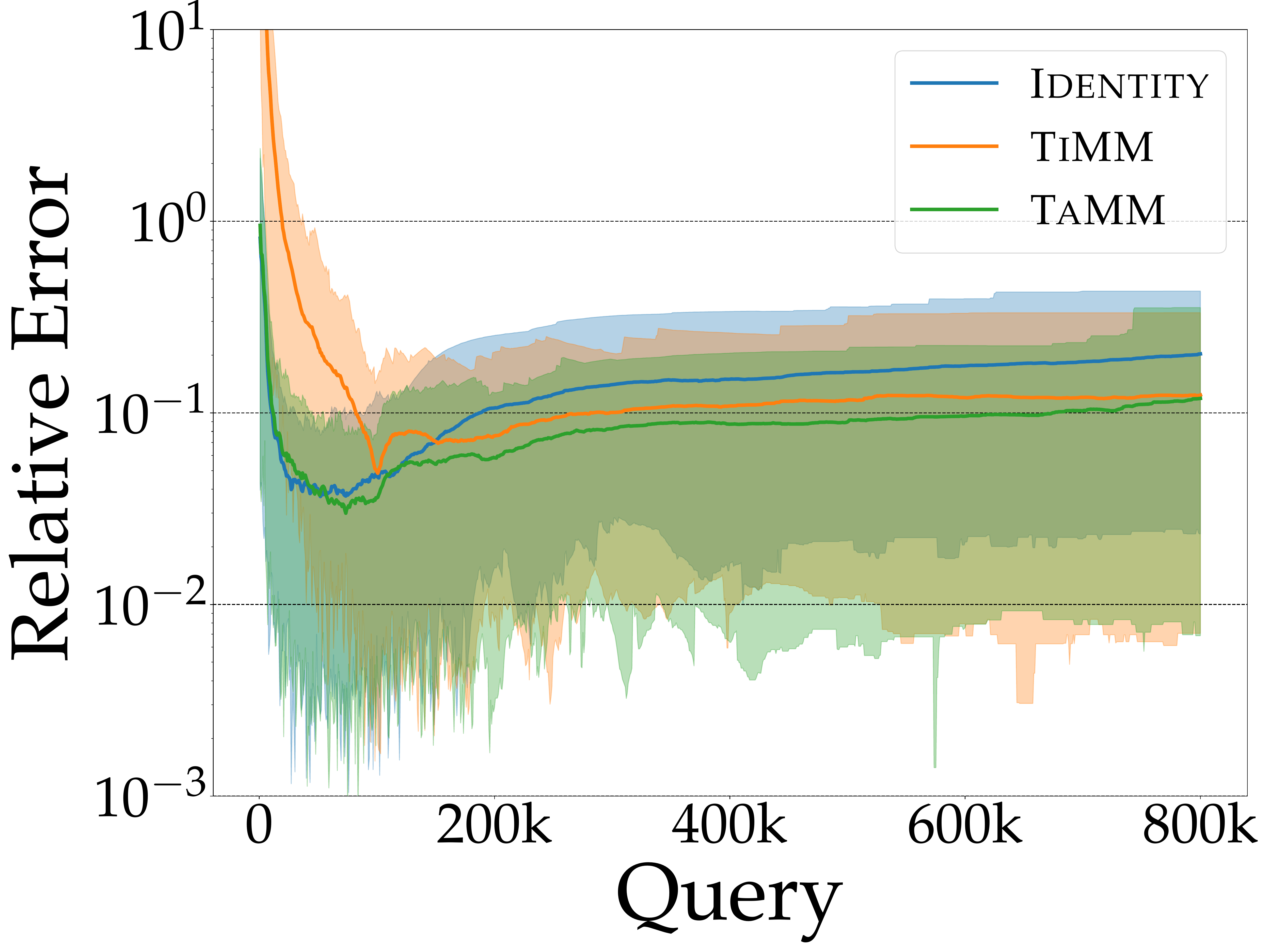}
        \caption{Truncated $\mm$}
        \label{fig:SVT_err_TMM}
    \end{subfigure}
    \begin{subfigure}[b]{0.5\linewidth}
    \centering
        \includegraphics[width=1\linewidth]{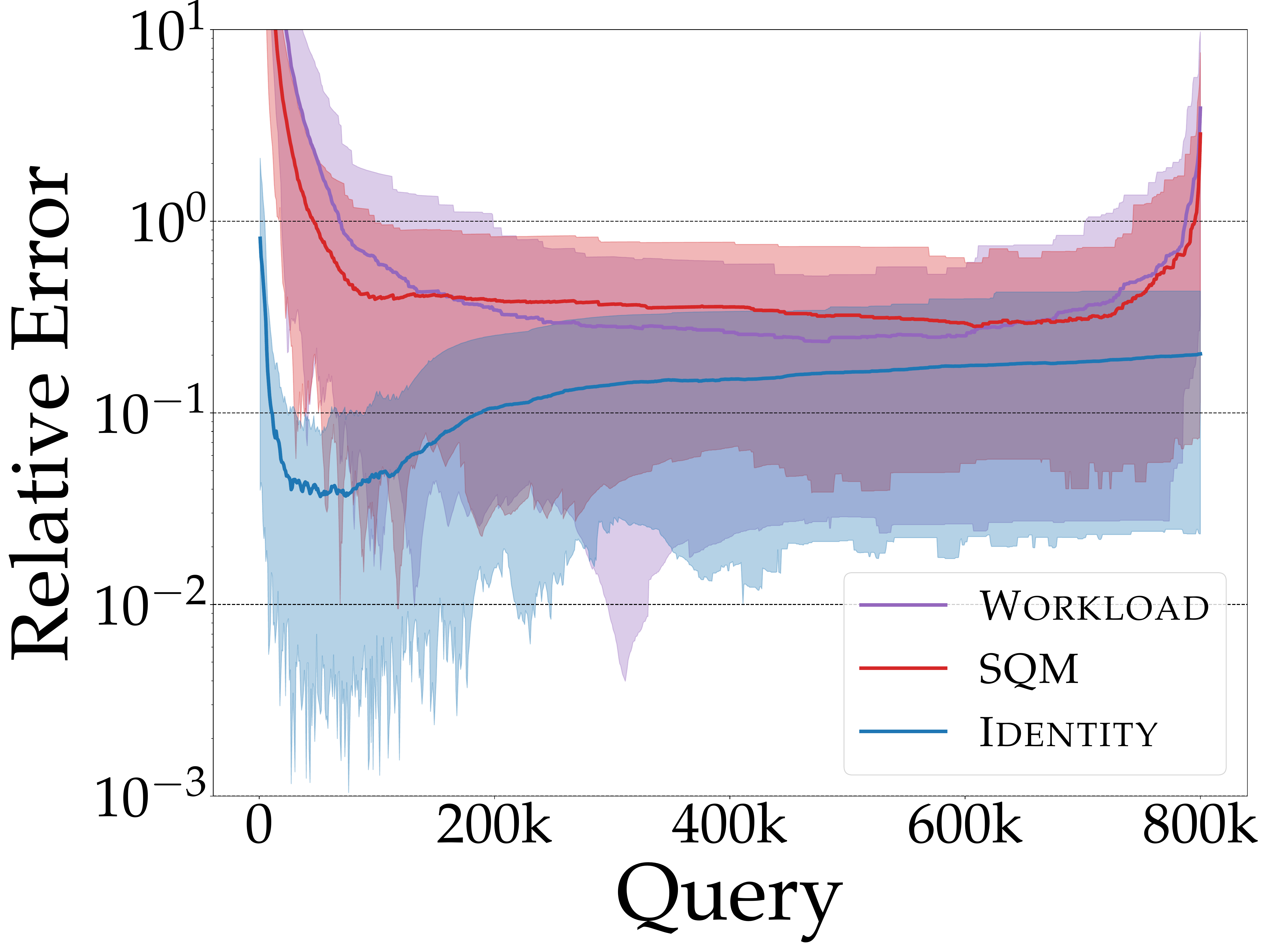}
        \caption{Baseline algorithms}
        \label{fig:SVT_err_wksqm}
    \end{subfigure}
    \caption{Relative Errors of Queries using $\ts$ in log scale}
    \label{fig:error}
\end{figure}

\begin{figure}[t]
  \begin{subfigure}[b]{0.5\linewidth}
    \centering
    \includegraphics[width=1\linewidth]{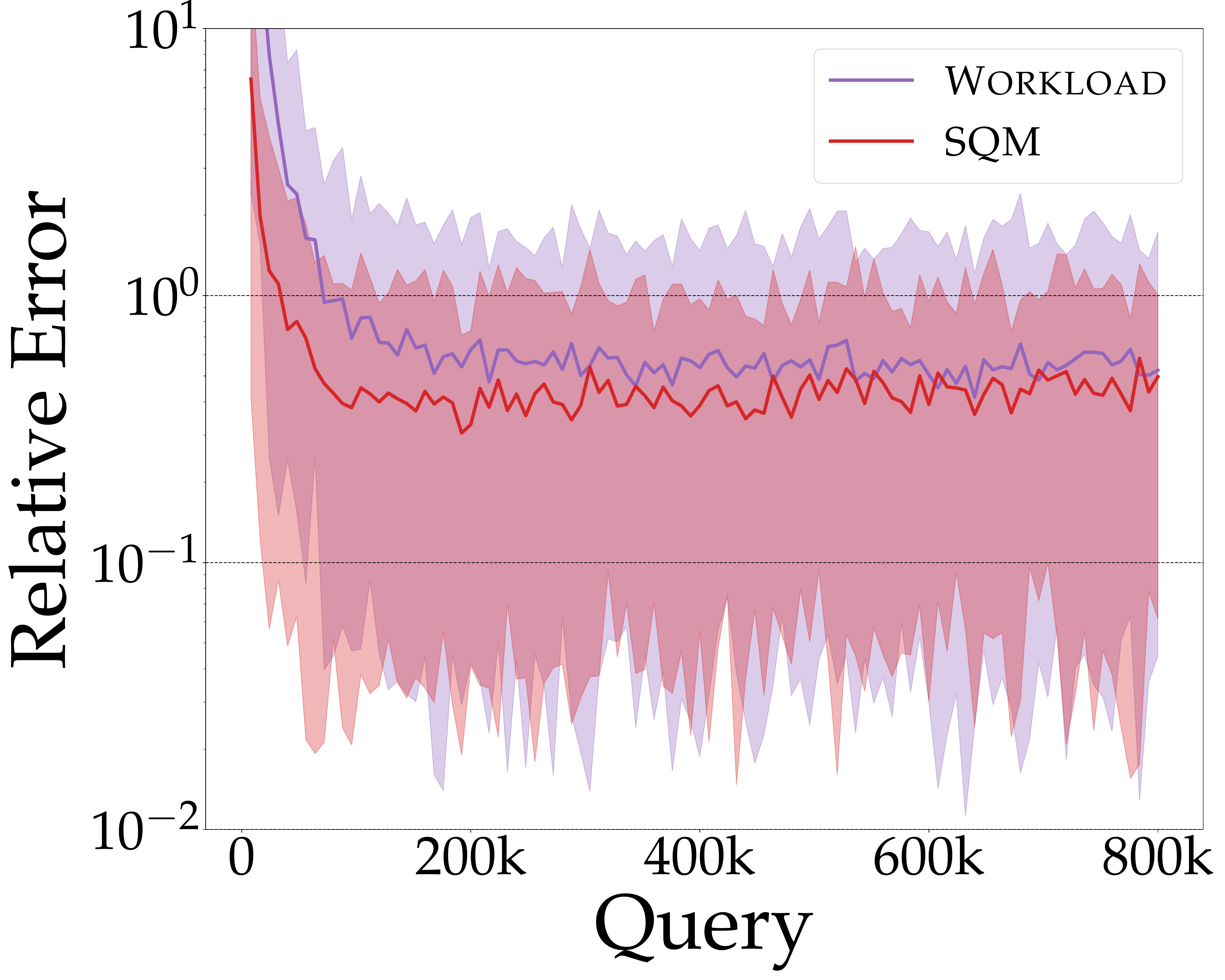}
    \caption{Without isotonic regression}
    \label{fig:noreg}
  \end{subfigure}
  \begin{subfigure}[b]{0.5\linewidth}
    \centering
    \includegraphics[width=1\linewidth]{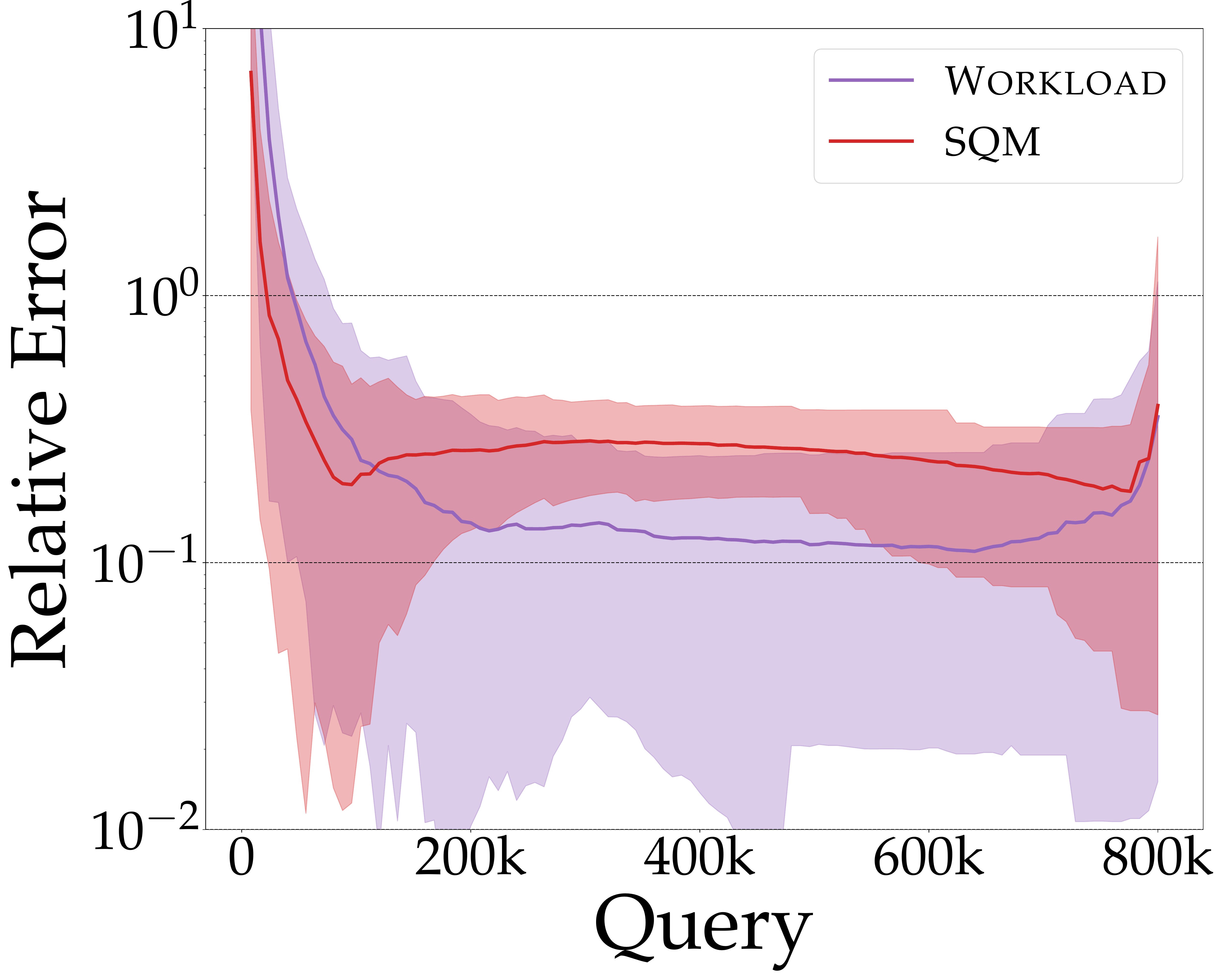}
    \caption{With isotonic regression}
    \label{fig:reg}
  \end{subfigure}
  \caption{Error of 100 queries using $\wk$ and $\SQM$ with $\SVT$ (log-scale)}
  \label{fig:100}
\end{figure}

\myparagraph{Results}
Our main experimental results are presented in \cref{fig:result,fig:error,fig:100}. Across all figures, the x-axis correspond prefix sum queries, for instance points at $x = 100k$ corresponds to query $q_{100k}$. In \cref{fig:result} the y-axis corresponds to the noisy answers, while in \cref{fig:error,fig:100} the y-axis shows the relative error. In \cref{fig:error} the solid black line corresponds to the true answers.
Solid colored lines represent the mean answer (or error) and the shaded areas cover 90\% (5 to 95 percentile) of the algorithm performance. Across all experiments, we observed that the $\id$ baseline dominated the rest of the baselines, this is expected due to the large size of the $Q_1$ workload. Thus, we mostly use $\id$ as the  baseline of comparison with our algorithms.

\myparagraph{Effects of Truncation}
    In \cref{fig:Identity} we compare the performance of the \algoname{Identity} algorithm with and without truncation. We can see that for $\ts$ and $\tr$ the overall variance of the noisy answers is much smaller than $\notrun$. This is expected as the scale of the Laplace noise added is significantly smaller when there is truncation -- especially so for larger queries. 
    Additionally and despite the fact that $\notrun$ is unbiased, the mean answers of $\notrun$ deviates from the true answers more than either $\ts$ or $\tr$.
    This is due to the large error of $\notrun$ and the bias caused by isotonic regression. Among the 2 different techniques to compute the truncation threshold, $\ts$ has smaller error. However, since both methods depends on several free parameters, we cannot say which one is better in general. In the following and for brevity, we present results using the $\ts$ technique.
    

\myparagraph{Truncated Matrix Mechanisms}
    In \cref{fig:SVT} we compare the answers from our new truncated matrix mechanisms with that of $\id$, the best performing baseline. We see that both $\jtm$ and $\itm$ offer less variance in their noisy answers than $\id$,  while their mean answers are comparable. To further examine the performance of these algorithms, in
     \cref{fig:error} we also present their relative errors.
    Overall, we see that $\jtm$ performs best across most queries and as the query size increases $\itm$ becomes more competitive.
    For very small queries the error of $\itm$ is approximately one order of magnitude larger than $\jtm$ and $\id$.
    This is due to $\itm$ adding noise to the weighted counts $\bm{Tx}$, while both $\id$ and $\jtm$ add noise to the counts $\bm{x}$.
    Thus, the noise of $\id$ and $\jtm$ is proportional to the weight of the bucket while the noise of $\itm$ is more evenly distributed.
    This makes the noise of $\itm$ significantly higher for small queries but similar to $\jtm$ for larger queries. This  also explains how $\itm$ has smaller error than $\jtm$ for very large queries.
    %
    %

\myparagraph{Effects of Isotonic Regression}
In \cref{fig:100} we present the performance of algorithms $\wk$ and $\SQM$ for answering workload $Q_2$ with (\cref{fig:reg}) and without (\cref{fig:noreg})  isotonic regression.  The main finding is that isotonic regression offers a bigger boost in terms of mean error on $\wk$ than in $\SQM$ and for the majority of queries the variance of errors of $\SQM$ is \textit{worsened}. More specifically, \cref{fig:reg} shows that the 5 percentile errors of $\SQM$ are significantly raised after applying isotonic regression, while its mean error is only slightly improved. 


As a reminder, $\wk$ uses the same truncation threshold $\theta$ across all queries of $Q_2$, while $\SQM$ has a different threshold $\theta_i$ for each query. This results in $\wk$ adding the same bias in each noisy answer and $\SQM$ adding different bias for each query. 
Due to our findings, we conjecture that isotonic regression and L2 minimization in general may increase error when noisy answers have different levels of bias in them.





\section{Future work}
In future work we hope to both test the algorithms proposed in other settings as well as develop more sophisticated algorithms.
Although the proposed algorithms work on a range of high sensitivity queries, they are only tested on sum queries. Likewise all the experiments used diagonal truncation matrices due to the nature of the prefix sums. Further analysis of the affects on other types of queries and non-diagonal weight matrices would be valuable. 

In \cref{sec:batch_queries} we introduced the idea of optimizing a strategy and truncation matrix jointly. The algorithms proposed do not reach this ideal and instead use a heuristic approach. As such developing an algorithm which optimizes both the strategy and truncation matrix jointly remains an open problem.

We saw that some post-processing techniques may worsen the performance of some algorithms, particularly when the algorithm introduces different bias to each answer. It is an interesting open problem to investigate.

\clearpage

\printbibliography

\clearpage
\appendix

\newcommand{\SVTs}{\textsc{SQM}_{\textsc{SVT}}}
\newcommand{\recs}{\textsc{SQM}_{\textsc{Recur}}}

\section{Algorithms}
\subsection{Single Query Mode}
\Cref{alg:notrun}, \cref{alg:SVTs} and \cref{alg:recs} are different algorithms used for the Single Query Mode ($\SQM$).\par
Let $D$ be a dataset in multi-set , $W = \{w_1, \dotsc\}$ is a query workload on $D$, where each $w_i$ is the prefix query of income sums at most $i$.

\begin{algorithm}[htp]
	\caption{\notrun$(D, W, \epsilon)$}\label{algo:no_trunc}	
\begin{algorithmic}
\State $\epsilon' \leftarrow \epsilon / |W|$
\State $ A \leftarrow \emptyset$
\ForAll{$w_i \in W$}
	\State  $ A \leftarrow \cup w_i(D) + \lap{i/\epsilon'}$
\EndFor
\Return $A$
\end{algorithmic}
\label{alg:notrun}
\end{algorithm}

$\rho$ is the ratio of splitting the privacy budget, $f_j$ is the counting query with uppber bound $j$, $c$ is the rate of increase of the counting query bound, $r$ is the ratio of the dataset set to be kept, $T$ is the threshold for the SVT algorithm. Truncate is the function to truncate $D$ by changing all rows less than $t$ by $t$.

\begin{algorithm}[htp]
    \caption{$\SVTs(D, W, \epsilon, \rho, c, s, r)$}\label{algo:sqm_svt}	
\begin{algorithmic}
\State $T \gets r|D|$
\State $\{f_j\} \gets \{f_s, f_{cs}, f_{c^2s}, \ldots \}$
\ForAll{$w_i \in W$}
	\State $t \gets \SVT(D,\{f_i\},T,\rho\epsilon)$
	\If{$t < i$}
		\State $D' \gets \textsc{Truncate}(D, t)$
		\State \textbf{Return} $w_i(D') + \lap{t/(1-\rho)\epsilon}$
	\Else
		\State \textbf{Return} $w_i(D) + \lap{i/(1-\rho)\epsilon}$
	\EndIf
\EndFor
\end{algorithmic}
\label{alg:SVTs}
\end{algorithm}

\begin{algorithm}[htp]
    \caption{$\recs(D, W, \epsilon, \rho, \theta)$}\label{algo:sqm_rec}	
\begin{algorithmic}
\State $\beta \gets 2\rho\epsilon/5$
\State $\mu \gets 0.5$
\ForAll{$w_i \in W$}
	\State $t \gets \rec(D,\beta, \theta, \mu, \rho\epsilon)$
	\If{$t < i$}
		\State $D' \gets \textsc{Truncate}(D, t)$
		\State \textbf{Return} $w_i(D') + \lap{t/(1-\rho)\epsilon}$
	\Else
		\State \textbf{Return} $w_i(D) + \lap{i/(1-\rho)\epsilon}$
	\EndIf
\EndFor
\end{algorithmic}
\label{alg:recs}
\end{algorithm}

\subsection{Batch Query Mode}
We call all other algorithms we use the Batch Query Mode ($\BQM$), where we operate on the workload matrix and the vector form of the data. \Cref{alg:vect} is used for vectorization and truncation, which is a common procedure for all $\BQM$ algorithms. \Cref{alg:id} and \cref{alg:wk} are baseline algorithms used together with $\SQM$. 2 new algorithms are \cref{alg:itm} and \cref{alg:jtm}. We implement all 4 algorithms together in our experiments using \cref{alg:bqm}.\par
We use the function \textsc{Threshold}$(D, \epsilon)$ to denote the selection of truncation threshold $t$. The algorithm can be SVT or recursive mechanism. We tested both in our experiments. We will omit the parameters of the truncation algorithms as they are listed in the section of single query mode.\par 
$\bm{x}$ is a vectorization of $D$ and $v$ is the vector of corresponding attribute. (i.e $A(\bm{x}_i) = \bm{v}_i$. $\bm{W}$ is the workload matrix for this vectorization. $\bm{V}$ is a diagonal matrix with $\bm{V}_{ii} = v_i$\par
\begin{algorithm}[htp]
    \caption{\textsc{Vectorization, Truncation}($D, W, n, \epsilon_1$)}
\begin{algorithmic}

	\State $\bm{x}, \bm{v} \gets \vect(D,n)$
	\State $\bm{W} \gets \vect(W,n)$
	\State $\bm{V} \gets \textsc{Diagonal}(\bm{v})$
	\State $t \gets \textsc{Threshold}(D, \epsilon)$
	\State $\bm{T} \gets \bm{V}$
	\ForAll{$\bm{V}_{ii} > t$}
		\State $\bm{T}_{ii} \gets t$
	\EndFor
	\State \textbf{Return} $\bm{W}, \bm{x}, \bm{V}, \bm{T}$
\end{algorithmic}
\label{alg:vect}
\end{algorithm}
We define $\bm{\tilde{b}}$ as a length $n$ vector with each entry an independent sample of Lap($1/\epsilon_2$) distribution. We use it as an input to replace $\epsilon$ for simplicity.\par
\begin{algorithm}[htp]
\caption{\id($\bm{W}, \bm{T}, \bm{x}, \bm{\tilde{b}}$)}
\begin{algorithmic}
	\State \textbf{Return} $\bm{WT}(\bm{x}+\bm{\tilde{b}})$
\end{algorithmic}
\label{alg:id}
\end{algorithm}
\begin{algorithm}[htp]
\caption{\wk($\bm{W}, \bm{T}, \bm{x}, \bm{\tilde{b}}$)}
\begin{algorithmic}
	\State \textbf{Return} $\bm{WTx}+ \|\bm{WT}\|_1\bm{\tilde{b}}$
\end{algorithmic}
\label{alg:wk}
\end{algorithm}
\textsc{GreedyH}$(\bm{W})$ is the function in ektelo, which returns a strategy matrix for the workload matrix $\bm{W}$. \textsc{LeastSquare}($\bm{A},\bm{z}$) returns the least square solution of $\bm{Ay} = \bm{\hat{x}}$.\par
\begin{algorithm}[htp]
\caption{\itm($\bm{W}, \bm{T}, \bm{x}, \bm{\tilde{b}}$)}
\begin{algorithmic}
	\State $\bm{A} \gets \textsc{GreedyH}(\bm{W})$
	\State $\bm{z} \gets \bm{ATx} + \|\bm{AT}\|_1\bm{\tilde{b}}$
	\State $\bm{\hat{x}} \gets \textsc{LeastSquare}(\bm{A},\bm{z})$
	\State \textbf{Return} $\bm{W\hat{x}}$
\end{algorithmic}
\label{alg:itm}
\end{algorithm}
\begin{algorithm}[htp]
\caption{\jtm($\bm{W}, \bm{T}, \bm{x}, \bm{\tilde{b}}$)}
\begin{algorithmic}
	\State $\bm{A} \gets \textsc{GreedyH}(\bm{WT})$
	\State $\bm{z} \gets \bm{Ax} + \|\bm{A}\|_1\bm{\tilde{b}}$
	\State $\bm{\hat{x}} \gets \textsc{LeastSquare}(\bm{A},\bm{z})$
	\State \textbf{Return} $\bm{WT\hat{x}}$
\end{algorithmic}
\label{alg:jtm}
\end{algorithm}
In the experiments, we share truncation threshold for all 4 mechanisms in one instance to control the effect of truncation. The algorithm used in the experiment is as follows.\par 
\begin{algorithm}[htp]
\caption{\BQM($D, W, n, \epsilon, \rho$)}
\begin{algorithmic}
	\State $\bm{W}, \bm{x}, \bm{V}, \bm{T} \gets \textsc{Vectorization, Truncation}(D, W, n, \rho\epsilon)$
	\State $\bm{\tilde{b}} \gets \lap{1/(1-\rho)\epsilon}_n$
	\State $y_1 \gets \id(\bm{W}, \bm{T}, \bm{x}, \bm{\tilde{b}})$
	\State $y_2 \gets \wk(\bm{W}, \bm{T}, \bm{x}, \bm{\tilde{b}})$
	\State $y_3 \gets \itm(\bm{W}, \bm{T}, \bm{x}, \bm{\tilde{b}})$
	\State $y_4 \gets \jtm(\bm{W}, \bm{T}, \bm{x}, \bm{\tilde{b}})$
	\State \textbf{Return} $y_1, y_2, y_3, y_4$
\end{algorithmic}
\label{alg:bqm}
\end{algorithm}

Detailed explanations and error analysis of the new proposed algorithms are below.
\subsection{Truncation-independent Matrix Mechanism}
In \cref{alg:itm} ($\itm$), the truncation matrix $\bm{T}$ is chosen without considering the strategy matrix. It can be chosen using truncation methods for single query as described for identity strategy.\par 
Then, we can choose a $m \times n$ strategy matrix $\bm{A}$ under the frame of matrix mechanism using workload matrix $\bm{W}$ and take $\bm{Tx}$ as the input. Thus, using Laplace mechanism, we have
\[ \cm{L}(\bm{A},\bm{Tx}) = \bm{ATx} + \frac{\Delta(\bm{AT})}{\epsilon}\bm{\tilde{b}} \]
where $\Delta(\bm{AT}) = \max_{j}\sum_{i=1}^m (\bm{AT})_{ij}$ is the sentivity of $\bm{AT}$, which is the maximum of the column sums of $\bm{AT}$, and $\bm{A}^+ = (\bm{A}^T\bm{A})^{-1}\bm{A}^T$ is the left pseudoinverse of $\bm{A}$.\par 
We then apply the matrix mechanism to have the output
\[ \cm{M}_{\bm{A}}(\bm{W},\bm{Tx}) = \bm{WA^+}\cm{L}(\bm{A},\bm{Tx}) = \bm{WTx} + \frac{\Delta(\bm{AT})}{\epsilon}\bm{WA^+\tilde{b}} \]
The error for one query $\bm{w}$ is
\begin{align*}
&\text{Error}[\cm{M}_{\bm{A}}(\bm{w},\bm{Tx})]\\ &= \E[(\bm{wDx}-\bm{wTx} - \frac{\Delta(\bm{AT})}{\epsilon}\bm{wA^+\tilde{b}})^2]\\
&= (\bm{w(D-T)x})^2 + \frac{2\Delta(\bm{AT})^2}{\epsilon^2}\bm{w}(\bm{A}^T\bm{A})^{-1}\bm{w}^T
\end{align*}
where $\E[(\bm{wA^+\tilde{b}})^2]$ is from the matrix mechanism.\par 
Thus, the total error is
\begin{align*}
&\text{Error}[\cm{M}_{\bm{A}}(\bm{W},\bm{Tx})]\\ &= \sum_{i=1}^p \text{Error}[\cm{M}_{\bm{A}}(\bm{w}_i,\bm{Tx})]\\ 
&= \|\bm{w(D-T)x}\|_2^2 + \frac{2\Delta(\bm{AT})^2}{\epsilon^2}\text{Tr}(\bm{W}(\bm{A}^T\bm{A})^{-1}\bm{W}^T)
\end{align*}
\subsection{Truncation-aware Matrix Mechanism}
\Cref{alg:jtm} ($\jtm$) uses matrix mechanism after truncation and the strategy matrix $\bm{A}$ is decided with the truncated weight matrix $\bm{T}$. Specifically, the matrix mechanism uses the weighted workload matrix $\bm{WT}$ as input.\par
We can implement a $m \times n$ strategy matrix $\bm{A}$ first and have 
\[ \cm{L}(\bm{A},\bm{x}) = \bm{Ax}+\frac{\Delta\bm{A}}{\epsilon}\bm{\tilde{b}} \]
where $\Delta \bm{A}=\max_{j}\sum_{i=1}^m (\bm{A})_{ij}$ is the sensitivity of $\bm{A}$.
Then, we can apply matrix mechanism by considering $\bm{WT}$ as the workload matrix and the result is thus
\begin{align*}
&\cm{T}_{\bm{A}, \bm{T}}(\bm{W}, \bm{x})\\ &= \bm{WTA^+} \cm{L}(\bm{A}, \bm{x})\\
&= \bm{WTA^+} (\bm{Ax} + \frac{\Delta \bm{A}}{\epsilon}\bm{\tilde{b}})\\
&= \bm{WT}(\bm{x}+ \frac{\Delta \bm{A}}{\epsilon}\bm{A^+} \bm{\tilde{b}}),
\end{align*} 
where $\bm{A}^+ = (\bm{A}^T\bm{A})^{-1}\bm{A}^T$ is the left pseudoinverse of $\bm{A}$.
In this case we have error for a single query $\bm{w}$ to be
\begin{align*}
&\text{Error}[\cm{T}_{\bm{A}, \bm{T}}(\bm{w}, \bm{x})]\\ &= \E[(\bm{wDx} - \cm{T}_{\bm{A},\bm{T}}(\bm{w}, \bm{x}))^2]\\
&= \E[(\bm{wDx} - \bm{wTx} - \frac{\Delta \bm{A}}{\epsilon}\bm{wTA^+\tilde{b}}))^2]\\
&=(\bm{w(D-T)x})^2 + \frac{2(\Delta \bm{A})^2}{\epsilon^2}\bm{wT}(\bm{A}^T\bm{A})^{-1}\bm{T}^T\bm{w}^T 
\end{align*}
as $\E[\bm{\tilde{b}}] = 0$ and $\E[(\bm{wTA^+ \tilde{b}})^2] = \bm{wT}(\bm{A}^T\bm{A})^{-1}\bm{T}^T\bm{w}^T$ from matrix mechanism.\par 
We thus have the total error as
\begin{align*}
&\text{Error}[\cm{T}_{\bm{A}, \bm{T}}(\bm{W}, \bm{x})]\\ &= \sum_{i=1}^p \text{Error}[\cm{T}_{\bm{A}, \bm{T}}(\bm{w}_i, \bm{x})]\\
&= \|\bm{W(D-T)x}\|_2^2  + \frac{2(\Delta \bm{A})^2}{\epsilon^2}\text{Tr}(\bm{WT}(\bm{A}^T\bm{A})^{-1}\bm{T}^T\bm{W}^T)
\end{align*}

\subsection{The equivalence of our truncation method and the second part of the recursive mechanism}
One important fact we used in our experiments is that for sum query specifically, the second part of recursive mechanism\cite{Chen13:recursive} is equivalent to the truncation mechanism we used, with $\hat{\Delta}$ in the recursive mechanism be considered as the truncation threshold $\theta$. We have the following theorem.
\begin{theorem}
When we consider $\hat{\Delta}$ as the truncation threshold $\theta$, and let $\epsilon_4 = \epsilon_2$ the second part of the recursive mechanism is equivalent to the truncation method.
\end{theorem}
The proof is as follows. As from the recursive mechanism 
\[ X = \min \{H_i + (N-i)\hat{\Delta}: 0\leq i \leq N \} \]
while $H_i$ is the sum of the lower $i$ weights. Thus, $H_i + (N-i)\hat{\Delta}$ is equal to the sum of a new data set with the lower $i$ weights unchanged and the $N-i$ weights left changed to $\hat{\Delta}$. Thus, the minimum of these sums will be keeping the weights smaller than or equal to $\hat{\Delta}$ unchanged and change weights larger than $\hat{\Delta}$ to $\hat{\Delta}$, exactly the same as truncation method using $\hat{\Delta}$ as the threshold.\par 
In addition, the noise added $Y_2 = \text{Lap}(\hat{\Delta}/\epsilon_4)$ is equal to the noise added in the truncation method $\text{Lap}(\theta/\epsilon_2)$ when $\epsilon_4 = \epsilon_2$. Thus, we can say the 2 methods are equivalent in the case of sum queries.

\end{document}